\documentstyle[aps,amsfonts]{revtex}
\input psfig.sty
\input epsf

\def\B{{\cal B}}
\def\C{{\cal C}}
\def\H{{\cal H}}

\def\S{{\cal S}}

\def\I{{\cal I}}
\def\L{{\cal L}}
\def\be{\begin{equation}}
\def\ee{\end{equation}}
\def\({{\biggl (}}
\def\){{\biggl )}}
\def\br{{\biggl \}}}
\def\bl{{\biggl \{}}
\def\la{{\biggl \langle}}
\def\ra{{\biggl \rangle}}

\begin{document}
\draft

\title{THERMODYNAMIC LIMIT FOR POLYDISPERSE FLUIDS\\}
\author{S. Banerjee\thanks{Electronic address: shubho@ipst.umd.edu}$^{1,2}$,
R. B. Griffiths$^1$, and M. Widom$^1$,}
\address{$^1$Department of Physics, Carnegie Mellon University,
Pittsburgh, Pa. 15213\\
$^2$Institute for Physical Science and Technology, University of Maryland,
College Park, Md. 20740~\thanks{Present address}}

\date{\today}
\maketitle

\begin{abstract}
We examine the thermodynamic limit of fluids of hard core particles
that are polydisperse in size and shape. In addition, particles may
interact magnetically.  Free energy of such systems is a random
variable because it depends on the choice of particles. We prove that
the thermodynamic limit exists with probability $1$, and is
independent of the choice of particles. Our proof applies to
polydisperse hard-sphere fluids, colloids and ferrofluids. The
existence of a thermodynamic limit implies system shape and size
independence of thermodynamic properties of a system.

\end{abstract}
\vspace{1cm}
Keywords: Thermodynamic limit, Free energy, Colloids, Random, Polydisperse,
hard-core, Ferrofluid, Polar fluid, Magnetic fluid 



\section{Introduction}

The free energy of a system of particles, as defined in statistical
mechanics, depends explicitly on the the size and shape of the
container holding the particles (the ``system shape'') through the
dependence of the partition function on these quantities.
Thermodynamics, in contrast, assumes a free energy density independent
of the system shape and size. For typical systems, as the system size
grows the contribution of the boundary to the statistical free energy
becomes negligible compared with the bulk contribution.  In the
thermodynamic limit (system size going to infinity) the free energy
density becomes independent of system shape and the boundary
conditions.

Ruelle~\cite{dr} and Fisher~\cite{mef} proved the existence of
thermodynamic limits for a large class of fluids and solids with
interactions that fall off faster than $r^{-3}$ at large separation.
Interactions which fall off as $r^{-3}$ or slower, complicate the
thermodynamic limit.  For neutral systems with coulombic interactions,
Lieb~\cite{lieb} proved the existence of a thermodynamic limit using
the screening of the interaction at large distances. For dipolar
interactions, Griffiths~\cite{rbg} proved the existence of a
thermodynamic limit for dipolar lattices using magnetization reversal
in a domain.  The present authors~\cite{banerjee} extended Griffiths'
proof to several non-lattice models with dipolar interactions.

The systems discussed above contain identical particles with specified
interactions. In this paper we study random systems, fluids
consisting of hard core particles that are polydisperse in size and
shape.  A proof of the thermodynamic limit for polydisperse fluids
is needed because they occur abundantly in nature and technology, and
because their thermodynamic properties are of great
interest~\cite{dickinson,smith,kofke,bartlett}.

For a system where the particles are chosen at random from a
distribution of size and shape, the free energy is a random variable
because it depends on the choice of particles. In general, proof of a
thermodynamic limit depends upon the subadditivity of free energy as
the system size grows. Thus we must exploit the theory of subadditive
random variables~\cite{akcoglu}.  Furthermore, depending on the size
and shape distribution of particles, a particular choice of particles
may or may not pack into the available volume. Due to these
complications, the proof requires a restriction on the particle size
and shape distribution in the form of a ``packing condition''.
Thermodynamic limits of random systems have been investigated
previously~\cite{lebowitz,vanenter,rosa}. Those results, however,
apply only to lattice models.

For our calculations, we fix the choice of particles in a canonical
ensemble.  Thus, we study ``quenched'' systems. In contrast, in an
``annealed'' system the relative concentration of particles of various
types are controlled using chemical
potentials~\cite{kofke,chayes}. The quenched and annealed free
energies have been studied extensively in the context of spin glass
systems~\cite{rosa,hemmen,koukiou,serva}.  The quenched free energy is
calculated by fixing the interaction between spins whereas the
annealed free energy is calculated by averaging the partition function
over the random interactions first. The quenched free energy is the
desirable quantity since in a real system the interactions are
fixed. However, it is often easier to calculate the annealed free
energy. It is claimed that for polydisperse fluids the annealed free
energy equals the quenched free energy~\cite{chayes}.

Polydisperse fluids often phase separate into different liquid or even
solid phases with different concentration of particle types in each
phase~\cite{dickinson,smith,kofke,bartlett}.  In our terminology
quenching fixes the overall concentration of various particle types
and not the particle concentration in each phase. This terminology
should not be confused with that often used in the
literature~\cite{bartlett} of polydisperse colloidal systems to
describe systems that have reached equilibrium (annealed) in relative
particle concentration between the two phases, and systems that have
not reached equilibrium (quenched).

In this paper we prove that the free energy density $F/V$ of any
random choice of particles goes to some finite limiting value ${f}$
with probability $1$, independent of the choice of particles, as the
system size goes to infinity.  In Sec.~\ref{hardcore} we carry out
the proof for fluids of polydisperse hard core particles. In
Sec.~\ref{maghardcore} we extend the proof to fluids of hard core
particles with magnetic interactions. These systems may be liquids
such as ferrofluids~\cite{rr}. Finally, in Sec.~\ref{conclusions}
we summarize our results and discuss some models that are not
covered by our proof.

\section{Hard core particles polydisperse in shape and size}
\label{hardcore}
In this section we show that, for a fluid of hard core particles
polydisperse in size and shape contained within a box of any shape,
the free energy density goes to a limit with probability $1$ as we
take the system size to infinity.  Our proof involves three main
steps. First, we consider systems of particles contained within
cubical containers and show that the free energy density of systems
shaped as cubes goes to a limit with probability $1$ as the system
size goes to infinity. In the second step, we show that this limiting
free energy density is independent of the choice of particles. In the
the final step, we prove system shape independence of the limit by
showing that the limiting free energy density for a system of any
arbitrary shape is the same as that of a system shaped as a cube.

In Sec.~\ref{definition} we define our model that includes the
particle size and shape distribution, a choice of particles from the
distribution, the Hamiltonian, the partition function, and the free
energy. We impose an artificial upper bound on the free energy to deal
with choices of particles that do not fit inside given containers
without overlapping. This section also defines a packing condition on
the particle size and shape distribution which guarantees that with
high probability the particles fit within containers without
overlapping.

In Sec.~\ref{cubes} we consider systems shaped as cubes and show
that the free energy density satisfies the conditions of a subadditive
ergodic theorem~\cite{akcoglu} so that the free energy density of a
cube-shaped system goes to a limit for almost every choice of
particles as the cube size goes to infinity. To make use of the
subadditive ergodic theorem we define a lattice and divide space into
basic cubes defined on the lattice. A choice of particles identifies
some specific particles associated with each basic cube. Any box
defined on the lattice holds particles assigned to the basic cubes
inside the box. The lattice, the basic cubes, and the boxes, are a
device for assigning particles to containers in a random but
well-defined fashion and employing the subadditive ergodic theorems in
the literature.

In Sec.~\ref{choiceindep} we show that the variance of the free
energy density around its mean vanishes as the cube size goes to
infinity, proving that the limit is in fact independent of the
specific choice of particles. In Sec.~\ref{shape-indep} we prove
the system shape independence of the free energy density by filling up
an arbitrary shaped system with cubes and showing that the free energy
density of any arbitrary shape has the same limit as that of a cube.

\subsection{Definitions and assumptions}
\label{definition}

Let $X$ be a set of particle sizes and shapes (see Fig. 1) with some
probability measure $\mu_1$ defined on it. Consider a container $S$,
of volume $V_S$ and a choice ${\bf c}_{N_S}$ of $N_S$ hard core
particles contained in $S$, independently chosen from $X$, one at a
time. Denote the set of all such choices by $\C_{N_S}$. The
probability measure $\mu_{N_S}$ defined on $\C_{N_S}$ is the
$N_S$-fold product of the probability measure $\mu_1$ defined on $X$.
Let ${\bf c}$ denote a choice of an infinite sequence of particles and
$\C$ denote the set of all possible choices. The probability measure
$\mu$ defined on $\C$ is the infinite-fold product of $\mu_1$.

The particles can be placed anywhere inside $\S$ with any orientation
as long as the Hamiltonian $\H_\S=\H^{HC}_S$ is finite, where
\begin{equation}\label{hc}
\H^{HC}_S({\bf c}_{N_S}) = 
   \left\{ \begin{array}{ll}
         0  \mbox{~~~~~~~~~if no particles overlap each other or the 
boundaries of}~S \\ 
        +\infty  \mbox{~~~~~otherwise,}
              \end{array}
              \right.
\end{equation}
is the hard core repulsion between the particles. Although not
indicated explicitly, $\H^{HC}_S ({\bf c}_{N_S})$ depends on the
particle center of mass positions $\{{\bf r}_i\}$ and the particle
orientations $\{\Omega_i\}$. Define the partition function
\begin{equation}
\label{pf}
Z_S({\bf c}_{N_S}) = {{1}\over{\Omega^{N_S} N_S!}} \int_S \prod_{i=1}^{N_S}
d{\bf r}_i  d\Omega_i e^{-\H_S({\bf c}_{N_S})/k_B T},
\end{equation}
where $\Omega=4\pi$ is the integral of $d\Omega$ over all possible
orientations of a particle. The prefactor $N_S!$ is chosen to make the
free energy extensive.  This factor may be modified depending on the
details of the set $X$ to account for the entropy of
mixing (see appendix~\ref{appd}).  However, the factor
$N_S!$ suffices as long as we do not ``unmix'' the
particles~\cite{dickinson,mixing}.

Because the particle sizes and shapes are drawn at random, sometimes a
choice of particles may not fit into the box $S$ without
overlapping. Some choices that fit into $S$ may have insufficient room
to move. Such choices have large, physically unrealistic, free
energies.  We wish to remove these unphysical choices from
consideration. However, their exclusion causes mathematical
inconvenience, since the probability measure for choices would no
longer be the product of the single particle probability measure on
$X$. On the other hand, including such choices causes the partition
function to vanish for some ${\bf c}_{N_S}\in \C_{N_S}$, leading to
infinite free energies.

To handle these difficulties, we define an arbitrary threshold
$z_0^{N_S}$, for the partition function of a container $\S$. The
constant $z_0>0$ can be interpreted as a (geometric) mean free phase
space volume per particle. We define the free energy in the following
artificial manner,
\begin{equation}
\label{freenrg}
F_S({\bf c}_{N_S})=\left\{ \begin{array}{ll}
              -N_S k_B {\rm T} \ln z_0~~~~~~~~~~~~~~\mbox{if}~Z_S({\bf c}_{N_S})
< z_0^{N_S} \\ 
              -k_B {\rm T} \ln Z_S({\bf c}_{N_S})  \mbox{~~~~~~~~~~otherwise.}
              \end{array}
              \right.
\end{equation}
Provided that the particle size and shape distribution $X$ obeys the
packing condition discussed below, choices with $Z_S({\bf c}_{N_S})<
z_0^{N_S}$ occur sufficiently infrequently that the limiting free
energy density is independent of the arbitrary constant $z_0$ (see
Sec.~\ref{choiceindep}).

Consider a cube $\Gamma$ of volume $V_{\Gamma}$ containing $N_\Gamma$
particles such that $\rho_\Gamma=N_\Gamma/V_\Gamma$.  Let $P_{\Gamma}$
be the set of choices ${\bf c}_{N_\Gamma}\in\C_{N_\Gamma}$ for which
$Z_{\Gamma_\Gamma}({\bf c}_{N_\Gamma})\ge z_0^{N_\Gamma}$, where
$N_\Gamma$ is the number of particles contained in $\Gamma$. Formally,
\be
\label{packingset}
P_{\Gamma}\equiv \bl {\bf c}_{N_\Gamma}\in \C_{N_\Gamma}~|~Z_{\Gamma}({\bf c}_{N_\Gamma})
\ge z_0^{N_\Gamma}\br
\ee
is the set of choices that ``pack'' in $\Gamma$.  Let $\{\Gamma\}$ be
a sequence of cubes. Our packing condition on the particle size and
shape distribution $X$ requires that for any $z_0>0$ there exists a
``critical packing density'' $\rho^*(z_0)>0$, such that for any
density $\rho_\Gamma \le \rho < \rho^*$,
\be
\label{packing}
\lim_{V_\Gamma \rightarrow \infty} \mu_{N_\Gamma}(P_{\Gamma})=1.
\ee
Here $\mu_{N_\Gamma}$ is the probability measure associated with the
sample space $\C_{N_\Gamma}$. See appendix~\ref{appd} for examples of
some particle size and shape distributions that satisfy the packing
condition.

\subsection{Thermodynamic limit for rectangular and cubical boxes}
\label{cubes}
In this section we show that the free energy density of rectangular
boxes goes to a limit with probability $1$ (i.e. for almost every
${\bf c}\in \C$) as the system size goes to infinity. We assume that
the particle distribution $X$ obeys the packing condition
Eq.~(\ref{packing}) and that the particle density $\rho < \rho^*$. For
the proof we make use of a subadditive ergodic theorem (theorem $2.7$
in Ref.~\cite{akcoglu}) by Akcoglu and Krengel.

Consider a tiling of non-negative real space ${\mathbb R}_+^3$ by
``basic'' cubes with each side of length $l_0$. The vertices of the
basic cubes define a lattice $\L$ such that any point in $\L$ can be
written as $(n_1,n_2,n_3)l_0$, where $n_i \in {\mathbb Z}_+$
(nonnegative integers).  For vectors $a=(a_i)$ and $b=(b_i)$ in $\L$,
the half-open interval $[a,b)$ denotes the set $\{u|u=(u_i)\in\L, a_i
\le u_i <b_i\}$. Physically, it represents a rectangular box in
${\mathbb R}_+^3$. We call $a$ the ``base'' of the box. Let $\I$ be
the class of all such rectangular boxes. Denote the cube
$[w,w+2^k{d})$ by $\Gamma_k^w$, where $w$ is any vector in $\L$, ${d}$
is the vector $(l_0,l_0,l_0)$ and $k$ is any nonnegative integer. Any
box $I\in \I$ can be completely filled with the ``basic'' cubes
$\Gamma_0^w$, where $w$ runs over the intersection of $\L$ with $I$.

Let ${\bf c}$ be a choice (infinite sequence) of particles. Break up
${\bf c}$ into successive sequences of $N_0$ particles and put one
such sequence into each basic cube $\Gamma_0^w$ using a one-to-one map
from a one-dimensional rectangular array (of sequences of particles)
to a three dimensional rectangular array (of basic cubes). Denote the
segment of particles in $\Gamma_0^w$ by ${\bf c}^w_{N_0}$.  For any
vector $u \in \L$, define a transformation $\tau_{u}$ on $\C$ such
that ${\bf c}'=\tau_{u}({\bf c})$ is another choice in $\C$ with
${{\bf c}_{N_0}^w}'={\bf c}_{N_0}^{w+u}$ for every $w \in S$. Since
$\mu$ is a product measure, $\tau_{u}$ is a measure preserving
transformation on $\C$.

Consider a rectangular box $I\in\I$, of volume $V_I$. Let the choice
of particles assigned to $I$ be the union of choices ${\bf c}_{N_0}^w$
assigned to basic cubes $\Gamma_0^w$ contained in $I$ (see
Fig. 2). The number of particles in $I$ is $N_I=N_0 V_I/V_{\Gamma_0}$.
The subadditive ergodic theorem requires that for each choice ${\bf c}
\in \C$ the free energy $F_I$ satisfies the following three
conditions:

\noindent ~~~(i) {\bf Translation invariance}: For every box 
$I \in \I$ and every vector $u \in \L$ 
\be
\label{translation}
F_I \circ \tau_u =F_{I+u}.
\ee

\noindent ~~~(ii) {\bf Subadditivity}: For any  box $I \in \I$ composed 
of nonoverlapping boxes $I_1,...,I_n \in \I$
\begin{equation}
\label{subadditivity}
F_I \le \sum_{i=1}^n F_{I_i}.
\end{equation}

\noindent ~~~(iii) {\bf Lower bound}: For some constant $\omega_A$ and for 
all $I \in \I$
\begin{equation}
\label{lowerbound1}
{1\over V_I} F_I \ge \omega_A,
\end{equation}
~~~where $V_I$ is the volume of box $I$.

\noindent Then, the theorem states that for a  {\it sufficiently 
regular} sequence of rectangular boxes $\{I_k\}$, of increasing size,
the free energy density $F_{I_k}/V_{I_k}$ goes to a limit $f({\bf c})$
almost everywhere in $\C$ as $k \rightarrow \infty$. Note that this
limit may depend on the choice of the particles ${\bf c}$. A sequence
of rectangular boxes $\{I_k\}$ of increasing size is sufficiently
regular (as defined by Akcoglu and Krengel~\cite{akcoglu}) if there
exists another sequence of rectangular boxes $\{I'_k\}$ such that box
$I'_k$ fully covers the box $I_k$, the ratio $V_{I_k}/V_{I'_k}$ is
always greater than or equal to some fixed nonzero constant, and
$\lim_{k \rightarrow \infty}I'_k={\mathbb R}_+^3$.

From our definition of the transformation $\tau_u$, the free energy
trivially satisfies~(\ref{translation}). For any box $I \in \I$
composed of nonoverlapping boxes $I_i\in \I$, the partition functions
satisfy
\begin{equation}
Z_I \ge \prod_i Z_{I_i},
\end{equation}
because each particle has more room to move inside $I$ than an
individual box $I_i$ (see Ref.~\cite{mef} for the detailed argument).  
The free energy therefore satisfies
subadditivity~(\ref{subadditivity}). The free energy also satisfies
the lower bound~(\ref{lowerbound1}) with $\omega_A= {k_B {\rm T} \rho}
\ln (\rho)$ which is temperature times the ideal gas entropy per
particle.


Hence, the free energy satisfies all the conditions of Akcoglu and
Krengel's subadditive ergodic theorem. To apply this to the special
case of a cube, construct a sequence of cubes of increasing size
$\{\Gamma_k\}$ such that $V_{\Gamma_k}\rightarrow \infty$ as $k
\rightarrow \infty$ and the distance of the bases of the cubes from
the origin grows less rapidly than their size. This ensures that the
sequence is sufficiently regular in the sense of Akcoglu and
Krengel~\cite{akcoglu}.  Then, according to the theorem the free
energy density $f_{\Gamma_k}\equiv F_{\Gamma_k}/V_{\Gamma_k}$ goes to
some limit $f_{\Gamma}({\bf c})$, the free energy density in a cube,
for almost every ${\bf c} \in \C$.

It is interesting to note that the sequence of cubes $\{\Gamma_k\}$
can be chosen so that no cubes overlap with each other and thus have
no particles in common. This suggests that the limiting free energy
density $f_\Gamma({\bf c})$ should not depend on ${\bf c}$. We
demonstrate that fact in the following section.

\subsection{Choice independence of the limiting free energy density}
\label{choiceindep}
In this section we prove that the limiting free energy density
$f_{\Gamma}({\bf c})$ is independent of the choice of particles ${\bf
c}$. We show that the free energy density averaged over all choices
goes to a limit and its variance goes to zero as the cube size goes to
infinity. This implies that the limit is same for any choice with
probability $1$. Furthermore, the limiting value and its variance are
independent of $z_0$, the arbitrary constant introduced in defining
free energy.

Let $\langle f_{\Gamma_k}\rangle$ denote the free energy density of a
cube $\Gamma_k$ averaged over all choices ${\bf c}_{N_k} \in \C_{N_k}$
of $N_k$ particles in ${\Gamma_k}$. From
subadditivity~(\ref{subadditivity}) of the free energy it follows that
\be
\langle f_{\Gamma_{k+1}}\rangle \le \langle f_{\Gamma_{k}} \rangle.
\ee
The sequence of average free energy densities $\{\langle
f_{\Gamma_{k}} \rangle \}$ is monotonically decreasing and has a lower
bound~(\ref{lowerbound1}). The average free energy density therefore
goes to some limit ${\bar f_{\Gamma}}$ as $k \rightarrow \infty$.

To show that the variance of the free energy density of cubes vanishes
as the cube size goes to infinity, we construct a cube $\Gamma_{k+1}$
by putting together eight nonoverlapping cubes $\Gamma_k$. By using
subadditivity~(\ref{subadditivity}), the lower
bound~(\ref{lowerbound1}) and the existence of a limiting average free
energy density for cubes we show (see appendix~\ref{appe}) that for
any $\epsilon >0$ there exists a sufficiently large $k_0$ such that
for any $k>k_0$,
\be
\label{variance}
\la \(f_{\Gamma_{k+1}} - \langle f_{\Gamma_{k+1}}\rangle\)^2 \ra  \le {1 \over 8}
\la  \(f_{\Gamma_{k}} - \langle f_{\Gamma_{k}}\rangle\)^2 \ra+ \epsilon.
\ee
Iterating this procedure $n$ times we write
\be
\la \(f_{\Gamma_{k+n}} - \langle f_{\Gamma_{k+n}}\rangle\)^2 \ra  \le {1 \over 8^n}
\la  \(f_{\Gamma_{k}} - \langle f_{\Gamma_{k}}\rangle\)^2 \ra+ 
\sum_{j=0}^{n-1}{1 \over 8^j} 
\epsilon.
\ee
For a fixed value of $k$ there exists $n_0$ such that, for all
$n>n_0$, the first term on the right hand side in the inequality is
less than $\epsilon$. The second term on the right hand side
approaches $(8/7)\epsilon$ from below. Hence, for any given
$\epsilon$, there exist $k_0$ and $n_0$ so that for any $k>k_0$ and
$n>n_0$, the variance in the free energy density satisfies the upper
bound
\begin{equation}
\la  \(f_{\Gamma_{k+n}} - \langle f_{\Gamma_{k+n}}\rangle\)^2 \ra \le 
{15 \over 7} \epsilon.
\end{equation}

For a sufficiently large system the variance of the free energy
density becomes arbitrarily small. We previously
proved~(Sec. \ref{cubes}) that $f_{\Gamma_k}$ goes to a limit
$f_\Gamma({\bf c})$ for almost every choice ${\bf c}$ of particles.  A
vanishing variance implies (by using Chebyshev's
inequality~\cite{billingsley}) that the probability of
$f_{\Gamma}({\bf c})$ differing from ${\bar f_\Gamma}$ by more than
some arbitrarily small fixed amount, also vanishes. The values of
$f_\Gamma({\bf c})$ thus converge ``in probability'' to ${\bar
f_\Gamma}$.  The existence of the limit $f_\Gamma({\bf c})$ with
probability $1$ together with convergence in probability to ${\bar
f_\Gamma}$ implies that $f_\Gamma({\bf c})$ equals ${\bar f_\Gamma}$
with probability $1$.  In appendix~\ref{appf} we show that ${\bar
f_\Gamma}$ is a convex and continuous function of the density $\rho$.

Finally, we show that the limit ${\bar f_\Gamma}$ and the variance are
independent of the arbitrary constant $z_0$. Let $P_{\Gamma_k}$ be the
set of choices that pack in a cube $\Gamma_k$ as defined in
Sec.~\ref{definition}. Let $Q_{\Gamma_k}$ be the complement of
$P_{\Gamma_k}$ so that
$\mu_{N_k}(P_{\Gamma_k})+\mu_{N_k}(Q_{\Gamma_k})=1$.  Write the free
energy density for a cube $\Gamma_k$ averaged over all choices as
\be
{\bar f_{\Gamma_k}}=\int_{P_{\Gamma_k}} d\mu_{N_k} ~f_{\Gamma_k}+
\int_{Q_{\Gamma_k}} d\mu_{N_k} ~f_{\Gamma_k}.
\ee
From the definition of free energy~(\ref{freenrg}) it follows that
\be
{\bar f_{\Gamma_k}}=\int_{P_{\Gamma_k}} d\mu_{N_k} ~f_{\Gamma_k}~-~k_B {\rm T} 
\ln z_0 ~\mu_{N_k}(Q_{\Gamma_k}).
\ee
By the packing condition,  $\mu_{N_k}(P_{\Gamma_k})\rightarrow 1$ and 
$\mu_{N_k}(Q_{\Gamma_k})\rightarrow 0$, 
as $k \rightarrow \infty$. Hence ${\bar f_{\Gamma}}$ is independent of $z_0$.
Now write the variance of the free energy density as
\be
\langle (f_{\Gamma_k}-{\bar f_{\Gamma_k}})^2 \rangle=\int_{P_{\Gamma_k}} 
d\mu_{N_k}
~(f_{\Gamma_k}-{\bar f_{\Gamma_k}})^2+\int_{Q_{\Gamma_k}} d\mu_{N_k}
~(f_{\Gamma_k}-{\bar f_{\Gamma_k}})^2.
\ee
From the definition of free energy~(\ref{freenrg}) it follows that
\be
\langle (f_{\Gamma_k}-{\bar f_{\Gamma_k}})^2 \rangle=\int_{P_{\Gamma_k}} 
d\mu_{N_k}
~(f_{\Gamma_k}-{\bar f_{\Gamma_k}})^2+
(k_B {\rm T} \ln z_0+{\bar f_{\Gamma_k}})^2~\mu_{N_k}(Q_{\Gamma_k}).
\ee
Since $\mu_{N_k}(Q_{\Gamma_k}) \rightarrow 0$ as $k \rightarrow
\infty$, because of the packing condition~(\ref{packing}), the
contribution of the choices ${\bf c}_{N_k} \in Q_{\Gamma_k}$ to the
variance vanishes. Hence the variance is independent of $z_0$. We can
then remove the free energy threshold in Eq.~(\ref{freenrg}) by taking
the limit of $z_0 \rightarrow 0$.

\subsection{System shape independence of the limiting free energy density}
\label{shape-indep}
Now we prove that for a sufficiently regular (in the sense of
Fisher~\cite{mef}) sequence of shapes $\{S_j\}$, the free energy
density goes to a limit as $V_{S_j}\rightarrow \infty$, provided that
the density of particles $\rho=N_{S_j}/V_{S_j}$ remains fixed. Note
that the volumes $V_{S_j}$ must be adjusted so that $N_{S_j}$ is an
integer.  To prove shape independence we show that the limiting free
energy density satisfies the bounds
\be
\label{ubound}
\lim_{j\rightarrow \infty} {\rm sup}~f_{S_j} \le {\bar f_\Gamma}
\ee
and
\be
\label{lbound}
\lim_{j\rightarrow \infty} {\rm inf}~f_{S_j} \ge {\bar f_\Gamma}
\ee
with probability $1$.  To show these bounds we use a technique similar
to that of Fisher in his proof for identical particles.

Consider a maximal filling of the shape $S_j$ by $n_{jk}$ cubes
$\Gamma_{k}^w$ so that each of the cubes $\Gamma_{k}^w$ is fully
contained within $S_j$ (see Fig. 3).  The length $l_0$ of the side of
the basic cubes $\Gamma_0$ is chosen such that there exists an integer
$N_0$ such that $N_0/l_0^3=\rho$. This ensures that it is possible to
achieve density $\rho$ in $\Gamma_{k}^w$ exactly. The total number of
particles in $S_j$ must equal $N_{S_j}=\rho V_{S_j}$. Therefore, in
general, not all $n_{jk}$ cubes contain equal numbers of
particles. Let there be two types of cubes, $A$ and $B$, such that
type $A$ cubes contain $N_{jk}=[N_{S_j}/n_{jk}]$ particles and type
$B$ cubes contain $N_{jk}+1$ particles independently chosen from the
particle distribution $X$. Here $[y]$ denotes the greatest integer
less than or equal to $y$. Note that because of our choice of $l_0$,
$N_{jk}$ must equal $2^k N_0$ plus some nonnegative integer. To ensure
that the number of particles in $S_j$ is $N_{S_j}$, the numbers of
type $A$ and type $B$ cubes (respectively $n_{jk}^A$ and $n_{jk}^B$),
obey
\be
\label{numbers}
N_{S_j}=n_{jk}^A N_{jk}+n_{jk}^B( N_{jk} +1)
\ee
and $n_{jk}=n_{jk}^A+n_{jk}^B$. 

Let $\rho_{jk}^A=N_{jk}/V_{\Gamma_k}$ and
$\rho_{jk}^B=(N_{jk}+1)/V_{\Gamma_k}$ be the density of particles in
cubes of type $A$ and $B$ respectively, and $\phi_{jk}^A
=n_{jk}^A/n_{jk}$ and $\phi_{jk}^B =n_{jk}^B/n_{jk}$ be the fractions
of cubes of type $A$ and $B$ respectively. Define
$n_{jk}V_{\Gamma_k}/V_{S_j}=1-\zeta_{jk}$, so that $\zeta_{jk}$ is the
fraction of volume in $S_j$ unfilled by the cubes $\Gamma_{k}^w$. The
density $\rho$ of particles in $S_j$ is related to the densities of
particles in the cubes by
\be
\label{density}
\rho=(1-\zeta_{jk}) (\phi_{jk}^A \rho_{jk}^A+ \phi_{jk}^B \rho_{jk}^B).
\ee
The density $\rho_{jk}^A$ must always be greater than or equal to
$\rho$ because if $\rho_{jk}^A<\rho$, then $\rho_{jk}^B \le \rho$
(since type $B$ cubes have only one more particle than type $A$
cubes), and Eq.~(\ref{density}) cannot be satisfied for $\zeta_{jk}>0$.
In the event that $\zeta_{jk}=0$ we have $\phi_{jk}^A=1$ and the equality
$\rho_{jk}^A=\rho$ holds.

Let ${\bf c}$ be a choice (infinite sequence) of particles. Put the
first $N_{S_j}$ particles of the sequence into $S_j$ and denote them
by ${\bf c}_{S_j}$.  Split ${\bf c}_{S_j}$ into $n_{jk}^A$ successive
sequences of $N_{jk}$ particles followed by $n_{jk}^B$ successive
sequences of $N_{jk}+1$ particles. Label these sequences as ${\bf
c}_{N_{jk}}^w$ and ${\bf c}_{N_{jk}+1}^w$ (depending on their length)
using a map from the one-dimensional array (of sequences of particles)
to a three-dimensional array (of cubes $\Gamma_{k}^w$). Put particles
sequences labeled by superscript $w$ into cube $\Gamma_{k}^w$.  From
the subadditivity of free energy it follows that
\be
F_{S_j}(\rho,{\bf c}_{S_j}) \le \sum_{w\in A} F_{\Gamma_{k}^w}
(\rho_{jk}^A,{\bf c}_{N_{jk}}^w)+
\sum_{w\in B} F_{\Gamma_{k}^w}(\rho_{jk}^B,{\bf c}_{N_{jk}+1}^w),
\ee
where the sum over $w\in A$ and $w\in B$ includes all cubes
$\Gamma_{k}^w$ of type $A$ and type $B$ respectively.  We include the
density as an argument for free energy and free energy density for the
sake of clarity where two or more different densities are present in
the same equation. The density is assumed to be $\rho$ if not
indicated explicitly.  The free energy densities satisfy
\be
f_{S_j}(\rho,{\bf c}_{S_j})~V_{S_j} \le V_{\Gamma_k}\bl{\sum_{w\in A} 
f_{\Gamma_{k}^w}
(\rho_{jk}^A,{\bf c}_{N_{jk}}^w)+\sum_{w\in B}f_{\Gamma_{k}^w}(\rho_{jk}^B,
{\bf c}_{N_{jk}+1}^w)\br}.
\ee
Rearranging terms we get 
\be
\label{free}
f_{S_j} (\rho,{\bf c}_{S_j})\le (1-\zeta_{jk}) \bl{{\phi_{jk}^A 
\over n_{jk}^A} 
\sum_{w\in A} 
f_{\Gamma_{k}^w}(\rho_{jk}^A,{\bf c}_{N_{jk}}^w)
+{{\phi_{jk}^B} \over n_{jk}^B} \sum_{w\in B} 
f_{\Gamma_{k}^w}(\rho_{jk}^B,{\bf c}_{N_{jk}+1}^w)\br}.
\ee
If $n_{jk}^B=0$, then the sum over type $B$ cubes on the right hand
side is not present because there are no type $B$ cubes.

Take the limit $j \rightarrow \infty$, holding $k$ fixed. For
sufficiently regular shapes (as defined by Fisher~\cite{mef})
$\zeta_{jk} \rightarrow 0$. Since $\rho_{jk}^A \ge \rho$ can only take
discrete values, it follows from Eq.~(\ref{density}) that there exists
$j_0$, such that for all $j \ge j_0$
\begin{eqnarray}
\rho_{jk}^A= \rho.
\end{eqnarray}
Therefore, $\phi_{jk}^A \rightarrow 1$ and $\phi_{jk}^B \rightarrow 0$
as $j \rightarrow \infty$.

The free energies $f_{\Gamma_k^w}(\rho)$ are identically distributed,
independent random variables, because the cubes $\Gamma_{k}^w$ are
identical in shape and the particles are independently chosen. By the
Strong Law of Large Numbers\cite{billingsley}
\be
\lim_{j \rightarrow \infty}
{1 \over n_{jk}^A} \sum_{w\in A} f_{\Gamma_k^w} \rightarrow 
\langle f_{\Gamma_k} \rangle
\ee
with probability 1 .
Therefore, the limit of Eq.~(\ref{free}) becomes
\be
\lim_{j\rightarrow \infty} {\rm sup}~f_{S_j} \le \langle f_{\Gamma_k}\rangle
\ee
with probability $1$. Now take the limit $k\rightarrow \infty$ to get the 
desired upper bound~(\ref{ubound}) with probability $1$.

To show the lower bound on the limiting free energy~(\ref{lbound})
enclose the shape $S_j$ completely inside the smallest possible cube
$\Gamma_K$ whose edge length is an integer multiple of $2^k l_0$ (see
Fig 3). Fill the empty space between $S_j$ and $\Gamma_K$ with
$n_{jk}$ cubes $\Gamma_k^w$ which lie completely outside $S_j$.
Fix the density of particles in
$\Gamma_K$ at $\rho$ by using two types of cubes $\Gamma_k^w$, type
$A$ cubes that contain $N_{jk}=[\rho (V_{\Gamma_K}-V_{S_j})/n_{jk}]$ particles
and type $B$ cubes that contain $N_{jk}^B=N_{jk}+1$ particles.

The number of particles in $\Gamma_K$,
\be
\label{numbers2}
N_{\Gamma_K}= N_{S_j}+n_{jk}^A N_{jk}+n_{jk}^B (N_{jk}+1),
\ee
where $n_{jk}^A$ and $n_{jk}^B$ are the number of type $A$ and type
$B$ cubes respectively.
Dividing by $V_{\Gamma_K}$ on both sides gives
\be
\label{density2v1}
\rho= {V_{S_j} \over V_{\Gamma_K}} \rho + 
\({{n_{jk} V_{\Gamma_k}}\over{V_{\Gamma_K}}}\)
(\phi_{jk}^A \rho_{jk}^A+\phi_{jk}^B \rho_{jk}^B).
\ee
We rewrite this relation, introducing $\Phi_j \equiv
V_{S_j}/V_{\Gamma_K} \le 1$, and introducing $\zeta_{jk} \ge 0$ as the ratio
of volume that is unfilled by the cubes $\Gamma_k^w$ to that of
$S_j$. Formally, $(V_{\Gamma_K}-n_{jk}V_{\Gamma_k})/V_{S_j}=1+\zeta_{jk}$.
Then Eq.~(\ref{density2v1}) becomes
\be
\label{density2}
\rho (1-\Phi_j) = 
(1-\Phi_j(1+\zeta_{jk})) (\phi_{jk}^A \rho_{jk}^A+\phi_{jk}^B \rho_{jk}^B).
\ee
It follows from Eq.~(\ref{density2}) that $\rho_{jk}^A \ge \rho$, because
if $\rho_{jk}^A < \rho$ then $\rho_{jk}^B \le \rho$ and Eq.~(\ref{density2})
cannot be satisfied for $\zeta_{jk} > 0$. In the event that $\zeta_{jk}=0$
then $\phi_{jk}^A=1$ and $\rho_{jk}^A=\rho$.

Let ${\bf c}$ be a choice (infinite sequence) of particles. Denote the
first $N_{\Gamma_K}$ particles of ${\bf c}$ by ${\bf
c}_{\Gamma_K}$. Split ${\bf c}_{\Gamma_K}$ into a subsequence of
$N_{S_j}$ particles followed by $n_{jk}^A$ successive subsequences of
$N_{jk}$ particles and $n_{jk}^B$ successive subsequences of
$N_{jk}+1$ particles. Denote the subsequence of $N_{S_j}$ particles by
${\bf c}_{S_j}$, and subsequences of $N_{jk}$ and $N_{jk}+1$ particles
by ${\bf c}_{N_{jk}}^w$ and ${\bf c}_{N_{jk}+1}^w$ respectively, using
a map from the one-dimensional array (of sequences of particles) to a
three-dimensional array (of cubes $\Gamma_{k}^w$).  Put particles
denoted by ${\bf c}_{S_j}$ into $S_j$ and particles labeled by
superscripts $w$ into cubes $\Gamma_{k}^w$.  From the subadditivity of
free energy it follows that
\be
F_{\Gamma_K}(\rho,{\bf c}_{\Gamma_K}) \le F_{S_j}(\rho,{\bf c}_{S_j})+
\sum_{w\in A} F_{\Gamma_{k}^w}(\rho_{jk}^A,{\bf c}_{N_{jk}}^w)+
\sum_{w\in B} F_{\Gamma_{k}^w}(\rho_{jk}^B,{\bf c}_{N_{jk}+1}^w),
\ee
Rewriting the above in terms of free energy densities and isolating
$f_{S_j}$ gives
\be
f_{S_j}(\rho,{\bf c}_{S_j}) \ge {V_{\Gamma_K} \over V_{S_j}} 
f_{\Gamma_K}(\rho,{\bf c}_{\Gamma_K})-
{V_{\Gamma_k} \over V_{S_j}} \bl{
\sum_{w \in A} f_{\Gamma_k^w}(\rho_{jk}^A,{\bf c}_{N_{jk}}^w) 
+\sum_{w \in B} f_{\Gamma_k^w}(\rho_{jk}^B,{\bf c}_{N_{jk}+1}^w)\br}.
\ee
Rearranging terms further, write
\begin{eqnarray}
f_{S_j} (\rho,{\bf c}_{S_j})\ge (1+\zeta_{jk}) \bl{{\phi_{jk}^A 
\over n_{jk}^A} 
\sum_{w\in A} 
f_{\Gamma_{k}^w}(\rho_{jk}^A,{\bf c}_{N_{jk}}^w)+{{\phi_{jk}^B} 
\over n_{jk}^B} 
\sum_{w\in B}
f_{\Gamma_{k}^w}(\rho_{jk}^B,{\bf c}_{N_{jk}+1}^w)\br}~~~~~~~~~~~~\\ \nonumber
~~~~~~~~~~~~~~+{V_{\Gamma_K} \over V_{S_j}} 
{ \bl{}f_{\Gamma_K}(\rho,{\bf c}_{\Gamma_K}) 
-\({\phi_{jk}^A \over n_{jk}^A} 
\sum_{w\in A} 
f_{\Gamma_{k}^w}(\rho_{jk}^A,{\bf c}_{N_{jk}}^w)+{{\phi_{jk}^B} 
\over n_{jk}^B} \sum_{w\in B} 
f_{\Gamma_{k}^w}(\rho_{jk}^B,{\bf c}_{N_{jk}+1}^w)\)\br}.
\end{eqnarray}

Take the limit $j \rightarrow \infty$ so that $\zeta_{jk} \rightarrow
0$. Since $\rho_{jk}^A\ge\rho$ is discrete, it follows from
Eq.~(\ref{density2}) that there exists $j_0$ such that
$\rho_{jk}^A=\rho$ for $j \ge j_0$. Therefore, $\phi_{jk}^A\rightarrow
1$ and $\phi_{jk}^B\rightarrow 0$ as $j
\rightarrow \infty$.  For the sequence $\{S_j\}$ of sufficiently
regular shapes~\cite{mef}, $ (V_{\Gamma_K}/V_{S_j}) \ge \delta_S$ for
some $\delta_S>0$. Using the Strong Law of Large
Numbers~\cite{billingsley} we get that the inequality
\be
\lim_{j\rightarrow \infty} {\rm inf}~f_{S_j} \ge \langle f_{\Gamma_k}\rangle
- {\delta_S} |f_\Gamma-\langle f_{\Gamma_k} \rangle|
\ee
is satisfied with probability $1$. Now take the limit 
$k\rightarrow \infty$ so that
$\langle f_{\Gamma_k} \rangle \rightarrow {\bar f_\Gamma}$. Therefore,
the lower bound~(\ref{lbound}) is satisfied with probability $1$.

Combining the two bounds~(\ref{ubound}) and~(\ref{lbound}) we get 
\be
\lim_{j\rightarrow \infty} f_{S_j} = {\bar f_\Gamma}
\ee
with probability $1$. Therefore, the limiting free energy density 
$f$ 
exists with probability $1$ for any shape and is independent of the shape 
of the system. Its value is equal to ${\bar f_\Gamma}$.

\section{Hard core magnetic particles polydisperse in size and shape}
\label{maghardcore}
In this section we extend our proof to fluids which consist of
polydisperse hard core particles, as in Sec.~\ref{hardcore}, but in
addition, they interact magnetically with each other. One example of
such a system is a ferrofluid~\cite{rr}, which is a colloidal
suspension of ferromagnetic particles in a carrier liquid. The
thermodynamic limit in this case is complicated by the $1/r^3$
fall-off of the magnetic interaction where the exponent of $1/r$
exactly matches the dimensionality of space.

We prove the existence of a thermodynamic limit with probability $1$
for fluids of magnetic hard core particles using a strategy similar to
that of hard core particles in Sec.~\ref{hardcore}. We show that the
free energy satisfies translation invariance~(\ref{translation}),
subadditivity~(\ref{subadditivity}), and the lower
bound~(\ref{lowerbound1}).  Once these relations are established, the
rest of the proof of a thermodynamic limit is identical to that of the
hard core particles in Sec.~\ref{hardcore} and is omitted to avoid
repetition.

\subsection{Definitions and assumptions}
Let the particles be chosen from a distribution of size, shape and
magnetization, $X$.  The magnetization is uniform inside each particle
and has a definite relation to the orientation of the particle. The
magnitude of the magnetization is the same for all particles. We
consider both superparamagnetic particles where the magnetization
${\bf M}_i$ can reverse independently of the particle
orientation~\cite{super1,super2} $\Omega_i$, and non-superparamagnetic
particles with magnetization fixed relative to particle
orientation. For the non-superparamagnetic particles, we require that
the particle shape have a two-fold rotation symmetry about any axis
perpendicular to its magnetization (see Fig. 4)~\cite{banerjee}.

Consider a container $S$ and a choice of particles ${\bf c}_{N_S}$.
The magnetic interaction between any two non-overlapping particles
$i,j$ inside $S$ is
\begin{equation}
\label{HM}
U_S^{ij}= \int_{v_i} d^3 {\bf r}~ \int_{v_j} d^3 {\bf r}'
{\biggl \{}{{\bf M}_i({\bf r}) \cdot {\bf M}_j({\bf r}') \over 
|{\bf r-r'}|^3}-
{3({\bf M}_i({\bf r}) \cdot ({{\bf r-r'}}))~({\bf M}_j({\bf r'}) \cdot 
({ {\bf r-r'}})) \over |{\bf r-r'}|^5}{\biggl \}},  
\end{equation}
where $v_i$ and $v_j$ are the regions of space occupied by the
magnetic material of particle $i$ and $j$, and ${\bf M}_i({\bf r})$ is
${\bf M}_i$ for ${\bf r}$ inside $v_i$ and zero outside. Write the
Hamiltonian as
\begin{equation}
\label{dhs}
\H_S({\bf c}_{N_S}) = 
   \left\{ \begin{array}{ll}
         \H_S^M  \mbox{~~~~~~if no particles overlap each other or the 
boundaries of}~S
\\ 
        +\infty   \mbox{~~~~~~otherwise,}
              \end{array}
              \right.
\end{equation}
where
\be
\H^M_S=\sum_{i<j=1}^N U^{ij}_S
\ee
is the magnetic interaction between the particles. 
Define the partition function $Z_S$ for the box in the same manner as
in~(\ref{pf}). For superparamagnetic particles where the magnetization
can rotate relative to the particle orientations, we include in
$\Omega_i$, in addition to the Euler angles, a discrete variable
$O_i=\pm 1$ specifying that the magnetization is parallel $(+1)$ or
opposite $(-1)$ to a direction fixed in the particle, and $\int
d\Omega_i$ includes a sum over $O_i$.  We may permit magnetization to
rotate independently of the particle orientation by augmenting the
integration in~(\ref{pf}) with an integration over the
direction of ${\bf M}$.  We require that the particle distribution
satisfy the packing condition~(\ref{packing}) and define the free
energy $F_S$ as in~(\ref{freenrg}).  The free energy, which depends on
the choice of particles, trivially satisfies the translation
invariance~(\ref{translation}).

\subsection{Subadditivity of the free energy}
\label{subfreenrg}
To show that the free energy satisfies
subadditivity~(\ref{subadditivity}), we consider a container $S$
composed of two adjacent nonoverlapping rectangular boxes $I_1,I_2 \in
\I$. Define the interaction energy between the two boxes $I_1, I_2$ by
\begin{equation}
\H_{I_1,I_2} \equiv \H_S-\H_{I_1}-\H_{I_2}
\end{equation}  
Let $F_S(\lambda)$ be the free energy of the combined system when the
Hamiltonian is $\H_{I_1}+\H_{I_2}$ plus a scaled interaction $\lambda
\H_{I_1,I_2}$.  Because $F_S(\lambda)$ is a concave function (that is,
$F_S''(\lambda)\le 0$), it is bounded above by
\begin{equation}
\label{expand}
F_S(\lambda) \le F_S(0)+\lambda F_S'(0),
\end{equation}
where the right side is a line tangent to the graph of $F_S(\lambda)$
at $\lambda=0$; here $F_S'(\lambda)$ and $F_S''(\lambda)$ are the
first and second derivatives.

As a consequence of~(\ref{expand}), the free energy $F_S(1)$ of the
fully interacting system satisfies the Gibbs inequality~\cite{falk}
\begin{equation}
\label{bogo}
F_S(1) \le F_S(0) + {[}\H_{I_1,I_2} {]}_{\lambda=0},
\end{equation}
where $F_S(0)=F_{I_1}+F_{I_2}$ is the free energy of the
non-interacting subsystems, and the classical ensemble average
\begin{eqnarray}
\label{pert}
{[}\H_{I_1,I_2} {]}_{\lambda=0}=~~~~~~~~~~~~~~~~~~~~~~~~~~~
~~~~~~~~~~~~~~~~~~~~~~~~~~~~~~~~~~~~~~~~~~~~~~~~~~~~~~~~~~~~~~~~~ \\ \nonumber
{{1}\over{\Omega^{N_{I_1}+N_{I_2}} N_{I_1}! N_{I_2}! Z_{I_1} Z_{I_2} }}
\int_{I_1} \prod_{i=1}^{N_1} d^3{\bf r}_i d\Omega_i
\int_{I_2} \prod_{j=1}^{N_2} d^3{\bf r}_j d\Omega_j 
\H_{I_1,I_2} e^{-(\H_{I_1}+\H_{I_2})/k_B {\rm T}}.
\end{eqnarray}

Consider a $\theta$ operator~\cite{banerjee} which acts on any given
system and has the following properties. It leaves the center of mass
positions ${\bf r}_i$ of particles unchanged, it maps the particle
orientations $\Omega_i$ on to themselves in such a way that it leaves
the integration measure $\prod_i d \Omega_i$ unchanged, and it leaves
the Hamiltonian $\H$ invariant.  In addition, it reverses the
magnetization of every particle.  For superparamagnetic particles,
spontaneous magnetization reversal acts as a $\theta$ operator.  For
non-superparamagnetic particles, a $180^\circ$ rotation of particles
about an axis of two-fold symmetry suffices as $\theta$ operator.

Applying the $\theta$ operator to $I_1$ but not $I_2$ leaves both
$\H_{I_1}$ and $\H_{I_2}$ invariant but reverses the sign of
$\H_{I_1,I_2}^M$ while leaving the integration measure
unchanged. Hence the average in~(\ref{pert}) vanishes and
from~(\ref{bogo}) we get
\be
\label{sub}
F_S \le F_{I_1}+F_{I_2}.
\ee
The free energy, therefore, satisfies subadditivity~(\ref{subadditivity}).

\subsection{Lower bound on the free energy}
The lower bound~(\ref{lowerbound1}) easily follows if the potential is
stable~\cite{dr,mef}. For stable potentials
\be
\label{stable}
\H_S \ge -\omega V_S,
\ee
where $\omega$ is some positive constant independent of the number of
particles.  Substituting the lower bound~(\ref{stable}) on $\H_S$
in~(\ref{pf}) gives an upper bound on the partition function and thus
a lower bound~(\ref{lowerbound1}) on the free energy.  Therefore, in
this section we will prove the stability for our model.

Let ${\bf M}_S({\bf r})$ be the magnetization distribution in a box
$S$, so that ${\bf M}_S({\bf r})={\bf M}_i$ if ${\bf r}$ is inside
particle $i$ and zero if ${\bf r}$ is outside any particle. Let ${\bf
H}^D_S({\bf r})$ be the magnetic field due to the particles in $S$. To
prove stability we make use of the positivity of field energy.  Adding
the magnetic self energy of each particle to $\H_S^M$ gives the total
energy of the whole system, considered as one magnetization
distribution (see Ref.~\cite{wfb} for identities on magnetization
distributions),
\begin{equation}
\label{prior}
\H^T_S=\H^M_S+\sum_{i=1}^{N_S} E_i^{self}=-{1\over{2}} \int d^3 {\bf r}~
{\bf H}_S^D({\bf r}) \cdot {\bf M}_S({\bf r}).
\end{equation}
Here ${\bf H}_S^D({\bf r})$ is the field, due to all particles inside
the container $S$ and
\begin{equation}
\label{selfnrg}
E_i^{self} = - {{1}\over{2}} {\bf M}_i \cdot \int_{v_i} 
d^3 {\bf r}~{\bf H}_{i}^D({\bf r}),
\end{equation}
where ${\bf H}_{i}^D({\bf r})$ is the field from magnetization ${\bf
M}_i$ of particle $i$ with volume $v_i$.

We place a lower bound on $\H$ by a method similar to that of
Griffiths~\cite{rbg}.  For any magnetization distribution ${\bf
M}({\bf r})$ and the field ${\bf H}^D({\bf r})$ caused by it
\begin{equation}
\label{identity}
-{1\over{2}} \int d^3 {\bf r}~{\bf H}^D({\bf r}) \cdot {\bf M}({\bf r})
={{1}\over{8 \pi}} \int d^3 {\bf r}~|{\bf H}^D({\bf r})|^2 \ge 0.
\end{equation}
Hence
\begin{equation}
\H^M_S+\sum_{i=1}^{N_S} E_i^{self} \ge 0.
\end{equation}
Brown~\cite{wfb} rewrites the self energy in~(\ref{selfnrg}) as
\begin{equation}
E_i^{self} = 2\pi \sum_{k,l} D_i^{kl} M_i^k M_i^l~~v_i,
\end{equation}
where ${\bf D}_i$ is the demagnetizing tensor of an ``equivalent
ellipsoid''; it exists for a particle of any shape~\cite{wfb}, and $k$
and $l$ index the components of ${\bf D}_i$ and ${\bf M}$. Since ${\bf
D}_i$ is positive definite, with trace equal to 1,
\begin{equation}
E_i^{self} \le 2\pi M^2 v_i,
\end{equation}
so that
\begin{equation}
\label{stability}
\H^M_S \ge -2 \pi M^2  \sum_{i=1}^{N_S} v_i.
\end{equation}
If the particles are unable to fit inside $S$ without overlap then
$\H_S =+\infty$ because of the hard core repulsion. Otherwise, if
particles fit inside $S$ without overlap, then $\sum_{i=1}^{N_S} v_i
\le V$ and
\be
\H_S=\H_S^M\ge -2\pi M^2 V_S.
\ee

The free energy of a box $I \in \I$ therefore satisfies the lower
bound~(\ref{lowerbound1}), and we have previously shown that it
satisfies the translation invariance~(\ref{translation}) and
subadditivity~(\ref{subadditivity}). The rest of the proof is
identical to that of hard core particles polydisperse in size and
shape and is omitted.

\section{Conclusion}
\label{conclusions}
We study the thermodynamic limit of hard core fluids consisting of
particles chosen from a distribution that is polydisperse in size and
shape. We show that a thermodynamic limit exists with probability $1$
and is independent of the choice of particles. The existence of a
thermodynamic limit implies shape {\it independence} of thermodynamic
properties. In this section we
discuss some models that are not covered by our proof, although
we believe they do possess thermodynamic limits.

This paper addressed fluids of particles with
hard core interactions and magnetic interactions. We believe the proof could
be extended to incorporate interactions that fall off faster than $1/r^3$.
One example is a
Lennard-Jones fluid with interaction
\be
U_{ij}={A_{ij} \over r^m} - {B_{ij}\over r^n}
\ee
between particles $i$ and $j$, where $m>n>3$ and $A_{ij}, B_{ij}>0$
are some constants chosen randomly for every pair of particles from
some distribution. A ``random'' Stockmayer fluid is another example,
where in addition to the Lennard-Jones interaction described above,
each particle $i$ has a randomly chosen magnetic dipole moment ${\bf
\mu}_i$.  Another example is a polydisperse charged colloid in ionic
solution~\cite{charged1,charged2} with an interaction such as
\be
U_{ij}={{q_i q_j e^{-\kappa r}}\over{\varepsilon r}}
\ee
where $\kappa>0$ is the Debye screening length.

For such systems the free energies for a container $S$ composed of two
adjacent nonoverlapping rectangular boxes $I_1,I_2 \in
\I$ separated by a distance $R\ge R_0>0$ satisfy
\be
\label{weak}
F_S \le F_{I_1}+F_{I_2}+\Delta_{12}.
\ee
Here
\be
\Delta_{12} = {N_{I_1} N_{I_2} \delta_{12} \over R^{3+\epsilon}}
\ee
with some $\epsilon>0$ and $\delta_{12}$ depending upon the
probability distributions of the pair potentials. Our proof of
thermodynamic limit uses a subadditive ergodic theorem which requires
strict subadditivity~(\ref{subadditivity}) with $\Delta_{12}=0$.
However, we suppose that provided that the probability distributions
are sufficiently restricted, a subadditive ergodic theorem similar to
that of Akcoglu and Krengel~\cite{akcoglu} but with the weaker
subadditivity condition~(\ref{weak}) will hold and a proof of a
thermodynamic limit will follow.

Granular magnetic solids provide other examples of systems outside the
realm of our present considerations. Consider a solid system with
particles frozen in space but magnetic moments able to rotate with
respect to the body of the particles~\cite{granular} (or a solid
containing rotationally free magnetic particles inside
cavities~\cite{ziolo}) One way to prepare such systems is to freeze
the carrier fluid in a colloidal suspension of magnetic hard core
particles so that the positions of particles are representative of an
equilibrium configuration of the fluid.  Our proof does not directly
apply to such systems because the infinite system cannot be
constructed from finite cubes of frozen particles. A system
constructed from finite cubes will have no particles overlapping the
boundaries of these cubes and thus the spatial arrangement of
particles differs from an equilibrium configuration of the fluid. If
the particles are ``unfrozen'' and frozen back again after their
positions have relaxed, then subadditivity~(\ref{subadditivity}) no
longer holds strictly, though it will hold on average.

\acknowledgements

We acknowledge useful discussions with A. Pisztora and L. Chayes. This work
was supported in part by NSF grant DMR-9732567 at Carnegie Mellon University
and by NSF grant CHE-9981772 at University of Maryland.
\appendix
\section{Examples of allowed particle distributions}
\label{appd}
Here we present some examples of particle size and shape distributions
which obey the packing condition. Our discussion hinges on the ability
to pack a monodisperse collection of hard spheres with a hexagonal
close packing fraction of ${\small \phi^{hcp}}={\pi \over 3 \sqrt 2}$.
Our presentation is informal and primarily intended to motivate the
form of our packing condition, rather than to achieve rigorous proofs.

{\bf Binary mixture of hard core spheres}: Consider spherical hard
cores of two types with radii $r_{\alpha}$ and $r_{\beta}$ occurring
with probabilities $p_\alpha$ and $p_{\beta}$, respectively.  A given
choice ${\bf c}_N \in {\cal C}_N$ of $N$ particles yields
$N_{\alpha}$ spheres of type $\alpha$ and $N_{\beta}=N-N_{\alpha}$ of type
$\beta$.
Define the partition function in a volume $V$ as
\begin{equation}
\label{pf2}
Z_V({\bf c}_N) \equiv {{1}\over{N_{\alpha}! N_{\beta}!}}
\int_V \prod_i dr_i e^{-\H_V({\bf c}_{N})/k_B T}.
\end{equation}
The factor of $N_{\alpha}! N_{\beta}!$ is chosen instead of $N!$
because it yields the usual entropy of mixing~\cite{mixing}.  Given a
choice ${\bf c}_N$, a volume $V({\bf c}_N)$ exists that is
sufficiently large that the partition function exceeds a threshold
\begin{equation}
\label{threshold}
Z_V({\bf c}_N) \ge z_0^N.
\end{equation}
We construct the volume $V({\bf c}_N)$ satisfying the
inequality~(\ref{threshold}) by assigning particles to spherical
shells of radius $r_{\alpha}'=r_{\alpha}+r_0$ and
$r_{\beta}'=r_{\beta}+r_0$ where ${{4\pi}\over{3}} r_0^3=z_0$. Shells
of type $\alpha$ and $\beta$ surround volumes
$v_{\alpha}'={{4\pi}\over{3}} r_{\alpha}'^3$ and
$v_{\beta}'={{4\pi}\over{3}} r_{\beta}'^3$.  If we choose the volume
\begin{equation}
V({\bf c}_N)=(N_{\alpha} v_{\alpha}' + N_{\beta} v_{\beta}')/
{\small \phi^{hcp}}
\end{equation}
then the inequality~(\ref{threshold}) holds.

To verify this claim, divide the volume into subregions of volumes
$N_{\alpha} v_{\alpha}'/{\small \phi^{hcp}}$ and $N_{\beta}
v_{\beta}'/{\small \phi^{hcp}}$ respectively. Closely pack the
$\alpha$- and $\beta$-type shells into their respective subregions and
evaluate the configurational integral in Eq.~(\ref{pf2}) in two
stages. First, confine each particle within its shell and integrate,
yielding a contribution of $z_0^N$. Second, permute identical
particles within identical shells, yielding a factor of $N_{\alpha}!
N_{\beta}!$. Since the combinatorial factor cancels against the
normalization of the partition function in Eq.~(\ref{pf2}), the
partition function $Z_V({\bf c}_N)$ is at least as large as $z_0^N$.

Now we show that, with probability 1, $V({\bf c}_N)$ does not greatly
exceed $N {\bar {v'}}(z_0)/{\small \phi^{hcp}}$, where we identify the
mean volume per shell ${\bar {v'}}(z_0) \equiv p_{\alpha} v_{\alpha}'
+ p_{\beta} v_{\beta}'$.  Note that ${\bar {v'}}(z_0)$ approaches the
mean particle volume ${\bar v}$ as $z_0 \rightarrow 0$. Take a cube
$\Gamma$ of volume
\begin{equation}
V_{\Gamma}={{N \bar{v'}}(z_0)\over{{\small \phi^{hcp}}}} (1 + \epsilon)
\end{equation}
where $\epsilon$ is an arbitrarily small positive number. Divide the
set ${\cal C}_N$ of choices ${\bf c}_N$ into a ``packing subset''
\begin{equation}
P_{\Gamma}=\{{\bf c}_N \in {\cal C}_N ~|~ V({\bf c}_N) \le V_{\Gamma}\}
\end{equation}
and a complementary ``nonpacking subset''
\begin{equation}
Q_{\Gamma}=\{{\bf c}_N \in {\cal C}_N ~|~ V({\bf c}_N) > V_{\Gamma}\}.
\end{equation}
Since $P_{\Gamma}$ and $Q_{\Gamma}$ are complementary subsets, their
measures obey $\mu_N(P_{\Gamma})+\mu_N(Q_{\Gamma})=1$. By Chebyshev's
inequality~\cite{billingsley} on the sum of independent random
variables we bound the measure of the nonpacking subset by
\begin{equation}
\mu_N(Q_{\Gamma})
\le {{p_{\alpha}^2 v_{\alpha}'^2 + p_{\beta}^2 v_{\beta}'^2}
\over{N {\bar {v'}}}(z_0)^2 \epsilon^2}.
\end{equation}
For sufficiently large $N$ the probability that a choice does not pack
into $V_{\Gamma}$ vanishes. Thus $V({\bf c}_N) \le V_{\Gamma}$ with
probability 1 at density
\begin{equation}
\rho \equiv N/V_{\Gamma} = {{{\small \phi^{hcp}}}\over{\bar{v'}(z_0) 
(1+\epsilon)}}.
\end{equation}
Therefore, binary mixture of hard spheres obeys the packing
condition~(\ref{packing}) with $\rho^*(z_0)={{{\small \phi^{hcp}}}
/{\bar {v'}}(z_0)}$.  Note that $\rho^*(z_0) {\bar v}$ approaches
${\small \phi^{hcp}}$ as $z_0 \rightarrow 0$.

{\bf Finitely many types of hard spheres}: Consider some finite
number, $t$, of types of hard spheres, each type with a different
finite radius and a certain probability of occurrence. The argument
presented above for a mixture of two types of spheres generalizes
easily to the case of $t>2$. The combinatorial factor in the partition
function~(\ref{pf2}) generalizes to $\Pi_\alpha N_\alpha!$ and the
mean volume per shell generalizes to ${\bar {v'}}(z_0) \equiv
\sum_\alpha p_\alpha v_\alpha'$. An argument similar to that for a 
binary mixture of hard spheres shows that this distribution satisfies
the packing condition~(\ref{packing}) with a critical packing density
$\rho^*(z_0)={{\small \phi^{hcp}} / {\bar {v'}}(z_0)}$.  A packing
fraction of ${\small \phi^{hcp}}$ is achievable in the limit $z_0
\rightarrow 0$, but for finitely many types of hard spheres a higher
packing fraction should be achievable because smaller particles can
occupy the interstitial sites of larger particles~\cite{polypack}.

The distributions above satisfy the packing condition using the
prefactor $\Pi_\alpha N_\alpha!$ to define the partition
function. However, the prefactor of $(\sum_\alpha N_\alpha)! = N!$ may
also be used to define the partition function as in
Eq.~(\ref{pf}). The logarithm of the ratio of the two prefactors is
the entropy of mixing between the particles of various types. As long
as the particles are chosen randomly from the distribution and not
``unmixed'', the factor of $N!$ suffices to give an extensive
entropy~\cite{dickinson,mixing}. To show that the distribution
satisfies the packing condition using prefactor $N!$, enclose
particles in shells so that they have free volume of at least $z_0'$
inside the shell. The partition function is bounded from below by
\be
Z_V \ge {\Pi_\alpha N_\alpha! \over N!} {{z_0'}^N} 
\ge \({z_0' \over t}\)^N.
\ee
Set $z_0'=t z_0$ to show that the distribution satisfies the packing
condition~(\ref{packing}) as long as $t$ is finite. Now the critical
packing density is $\rho^*(z_0)={{\small \phi^{hcp}} / {\bar {v'}}(t
z_0)}$.  As before, $\rho^*(z_0) {\bar v}$ approaches ${\small
\phi^{hcp}}$ as $z_0 \rightarrow 0$.

{\bf Continuous distribution of hard spheres with an upper size
cut-off}: Consider a continuous distribution of hard spheres with
volumes $0 \le v < v_{max}$.  This distribution has essentially an
infinite number of types of particles and so our discussion above does
not directly apply to it, however, it still satisfies the packing
condition~(\ref{packing}). To show this, define the partition function
with prefactor $N!$ to make the free energy extensive. Break the
distribution into a finite number, $t$, of bins with boundaries
$0<v_1<v_2...<v_t=v_{max}$. Let the $i^{th}$ bin $\B_i$ contain all
the particles with $v_{i-1} \le v < v_i$. Enclose every particle in
bin $\B_i$ within a shell of volume
\be
\label{shell-room}
v_{\B_i}' = {4 \pi \over 3} \bl \({3 v_i \over 4 \pi}\)^{1/3}+
\({3 z_0 \over 4 \pi}\)^{1/3} \br^3 
\ee 
so that the largest particle in $\B_i$ has a free volume of at least
$z_0'=t z_0$, then apply the same argument as for finitely many types
of spheres. The resulting $\rho^*(z_0)={{\small \phi^{hcp}} / {\bar
{v'}}(t z_0)}$ results in a suboptimal achievable packing fraction
$\phi=\rho^*(z_0) {\bar v}= {\small \phi^{hcp}} {\bar v}/ {\bar
{v'}}(t z_0)<{\small \phi^{hcp}}$ due to the bin width and the choice
of prefactor $N!$.  By using a sufficiently large but finite number of
bins and taking $z_0 \rightarrow 0$, a packing fraction arbitrarily
close to ${\small \phi^{hcp}}$ can be achieved.  As in the case of
discrete distributions, for a continuous distribution of hard spheres
one expects a higher packing fraction than that of a monodisperse
system should be achievable~\cite{polypack}.

{\bf Continuous distribution of hard spheres with no particle size
cut-off}: Now we consider a continuous distribution of hard core
spheres such that the particle size has no upper limit. To ensure that
the large sized particles occur sufficiently infrequently, we assume
that there exists a particle volume $v^*$ such that the probability of
a particle with volume greater than $v_0$ for any $v_0>v^*$ falls off
as
\be
\label{size-condition}
\mu_1(v>v_0) \le {\gamma \over v_0^{1+\delta}},
\ee
where $\gamma$ and $\delta$ are positive constants. Note that this
condition guarantees the existence of a mean particle volume but not
the existence of higher moments.

To show that this distribution satisfies the packing
condition~(\ref{packing}) we break the distribution into a finite
number, $t+1$, of bins with boundaries
$0<v_1<v_2...<v_{t-1}<v_t<{\infty}$.  We presume that $v_{t-1} >
v^*$. The next-to-last bin $\B_t$ includes all the particles with
$v_{t-1} \le v < v_t$ and the last bin $\B_{t+1}$ contains all the
particles with $v_t \le v < \infty$. We now send $v_t$ to infinity in
such a manner that $\B_{t+1}$ is empty with high probability but the
maximum particle size in $\B_t$ grows in a controlled fashion.  To
show that the distribution satisfies the packing
condition~(\ref{packing}) we first show that the set of choices with
no particle in $\B_{t+1}$ is a set of measure $1$. Then we show that
in this set there is a subset of measure $1$ of choices which have a
partition function of at least $z_0^N$.

The probability that a collection of $N$ particles randomly chosen
from the distribution contains at least one particle in the last bin
$\B_{t+1}$ is bounded by
\be
\mu_N({\bf c}_N~|~{\rm at~least~one~particle~in}~\B_{t+1}) 
\le {N \gamma \over v_t^{1+\delta}}.
\ee
We wish to avoid all choices with at least one particle in
$\B_{t+1}$. To ensure that the probability of such choices goes to
zero for large $N$, $v_t$ must grow faster than $N^{1/ (1+\delta)}$. At
the same time $v_t$ must grow slower than $N$ to ensure that the
largest particles in $\B_t$ are much smaller than the system
size. Hence let $v_t$ grow proportional to $N^{1/ (1+{\delta \over
2})}$, so that
\be
\mu_N({\bf c}_N~|~{\rm at~least~one~particle~in}~\B_{t+1}) \le 
{\gamma' \over N^{\delta \over 2+\delta}} \rightarrow 0,
\ee
where $\gamma'$ is some constant. For large $N$, the probability
approaches $1$ that no particle is in bin $\B_{t+1}$.

Now consider the particles in bin $\B_{t}$. The next-to-last bin
$\B_t$ must be treated specially because the largest particle size in
$\B_t$ diverges as the system size to goes infinity.  Enclose each
particle $j$ (with volume $v_j$) in $\B_t$ inside its own shell with
volume $v_j'$ so that it has a free volume of $z_0'=t z_0$.  Using
condition~(\ref{size-condition}) the total volume of the shells
corresponding to bin $\B_t$ (averaged over choices ${\bf c}_N$) obeys
\be
\la \sum_{v_j \in \B_t} v_j'\ra \le {\gamma (1+\delta) \over \delta} 
{N \over v_{t-1}^\delta}.
\ee
Since the volume of the system is proportional to $N$, the volume
fraction occupied by shells corresponding to bin $\B_t$ is
proportional to $v_{t-1}^{-\delta}$.  This volume fraction can be made
arbitrarily small by choosing a sufficiently large $v_{t-1}$. We
assume that the $N_t$ particles in bin $\B_t$ (enclosed in their
shells) can be packed inside a small region of the system in $N_t!$
different ways with high probability (tending to 1 as $N$ goes to
infinity).

The particles in the remaining bins $\B_1, ...,\B_{t-1}$ have a
partition function of at least $z_0^{N-N_t}$ in the remaining volume
of the system with high probability (tending to 1) as we showed
previously for a continuous distribution of spheres with an upper size
cut-off.  Therefore, these particles together with the particles in
$\B_{t}$ satisfy the packing condition~(\ref{packing}).

{\bf Non-spherical particles}: Finally, consider a distribution of
non-spherical particles.  To show that the distribution satisfies the
packing condition~(\ref{packing}), enclose each particle inside a
spherical shell with a diameter equal to the largest dimension of the
particle.  If the volume of the shells satisfy
condition~(\ref{size-condition}) then the rest of the argument is the
same as that of a continuous distribution of hard spheres.  Although
this argument shows the existence of a finite particle number density
$\rho^*(z_0)$, it does not guarantee the existence of a finite packing
fraction. In fact, it is possible to have distributions with oddly
shaped particles (fractals for example) so that a finite $\rho^*(z_0)$
exists but the particle packing fraction is zero.

\section{Variance of free energy density for cubes}
\label{appe}
To show the variance relation~(\ref{variance}) consider two adjacent 
nonoverlapping cubes $\Gamma_k^l$ and $\Gamma_k^m$ such that 
$I_k=\Gamma_k^l \cup \Gamma_k^m$ is a rectangular box. Using the subadditive 
relation~(\ref{subadditivity}) we get
\begin{equation}
\label{nrg-density}
f_{I_k} \le {1 \over 2} (f_{\Gamma_k^l}+f_{\Gamma_k^m}).
\end{equation}
Adding $\omega_A$ (see lower bound~(\ref{lowerbound1})) to both sides 
of~(\ref{nrg-density}) 
and squaring gives
\begin{equation}
f_{I_k}^2+2\omega_A f_{I_k}+\omega_A^2 \le {1 \over 4} (f_{\Gamma_k^l}
+ f_{\Gamma_k^m})^2+\omega_A^2+\omega_A (f_{\Gamma_k^l} + f_{\Gamma_k^m}).
\end{equation}
Averaging both sides over all choices of particles we write
\be
\langle f_{I_k}^2\rangle+2\omega_A \langle f_{I_k} \rangle+\omega_A^2 \le 
{1 \over 2} 
\langle f_{\Gamma_k}^2\rangle+{1 \over 2} {\langle f_{\Gamma_k}\rangle}^2+2 
\omega_A 
{\langle f_{\Gamma_k}\rangle}+\omega_A^2.
\end{equation}
Subtracting $\omega_A^2+\langle f_{I_k}^2 \rangle+2\omega_A \langle f_{I_k} 
\rangle$ 
and rearranging terms on the right-hand side  gives
\begin{equation}
\langle  (f_{I_k} - \langle f_{I_k} \rangle)^2 \rangle  \le {1 \over 2}
\langle  (f_{\Gamma_k} - \langle f_{\Gamma_k} \rangle)^2 \rangle+\langle 
f_{\Gamma_k}\rangle^2-
\langle f_{I_k}\rangle^2+2\omega_A(\langle f_{\Gamma_k} \rangle-
\langle f_{I_k} \rangle).
\end{equation}
Using the bound
\begin{equation}
\langle f_{\Gamma_k} \rangle+\langle f_{I_k} \rangle \le -2k_B {\rm T} 
\ln z_0 \equiv 2 \omega_B,
\end{equation}
we get 
\begin{equation}
\label{var}
\langle  (f_{I_k} - \langle f_{I_k} \rangle)^2 \rangle  \le {1 \over 2}
\langle  (f_{\Gamma_k} - \langle f_{\Gamma_k} \rangle)^2 \rangle+2(\omega_A+
\omega_B)
(\langle f_{\Gamma_k} \rangle-\langle f_{I_k} \rangle).
\end{equation}
Both ${\langle f_{I_k}\rangle}$ and $\langle f_{\Gamma_k} \rangle$ go to the 
same limit 
$f_\Gamma$ since 
${\langle f_{\Gamma_{k+1}}\rangle}\le\langle f_{I_k} \rangle\le
\langle f_{\Gamma_k} \rangle$. Therefore, we can bound the second term on the 
right in~(\ref{var}) below any $\epsilon' >0$ for sufficiently large $k$. 
It follows that
\begin{equation}
\label{var2}
\langle  (f_{I_k} - \langle f_{I_k} \rangle)^2 \rangle  \le {1 \over 2}
\langle  (f_{\Gamma_k} - \langle f_{\Gamma_k} \rangle)^2 \rangle+\epsilon'.
\end{equation}
By stacking two rectangular boxes $I_k$ together to form a square slab, 
then putting together 
two adjacent slabs, we construct a cube $\Gamma_{k+1}$. The variance of the cube 
$\Gamma_{k+1}$ is related to that of $\Gamma_{k}$ by
\begin{equation}
\label{var3}
\langle  (f_{\Gamma_{k+1}} - {\langle f_{\Gamma_{k+1}}\rangle})^2 \rangle  \le {1 \over 8}
\langle  (f_{\Gamma_k} - \langle f_{\Gamma_k} \rangle)^2 \rangle+ \epsilon,
\end{equation}
where $\epsilon=(1+{1\over 2}+{1\over 4})\epsilon'={7\over 4} \epsilon'$.

\section{Convexity  and continuity of the free energy density for cubes}
\label{appf}
Consider a cube $\Gamma_k$ with volume $V_{\Gamma_k}$ that contains
$N_k$ particles.  Since $N_k$ is an integer, the free energy density
$f_{\Gamma_k}(\rho,{\bf c}_{N_k})$ is defined only for discrete values
of $\rho$. Since $f_{\Gamma_k}(\rho,{\bf c}_{N_k})$ is a random
variable, a linear interpolation to include fractional number of
particles is not possible. However, the average free energy density
$\langle f_{\Gamma_k}(\rho)
\rangle$ is 
not random and we define $\langle f(\rho) \rangle$ for all values of
$\rho$ by linear interpolation between values of $\rho$ that
correspond to integer values of $N_k$.

Now consider a cube $\Gamma_{k+1}$ consisting of $4$ cubes $\Gamma_k$ each
containing $N_k'$ particles and $4$ cubes $\Gamma_k$ each
containing $N_k''$ particles. From subadditivity~(\ref{subadditivity})
of free energy, and averaging over all possible choices of particles in the 
cubes $\Gamma_k$, it follows that 
\be
\label{convex}
\la f_{\Gamma_{k+1}}(\rho) \ra \le {1\over 2}\(
\la f_{\Gamma_k}(\rho')\ra+\la f_{\Gamma_k}(\rho'')\ra\),
\ee
where $\rho'=N_k'/V_{\Gamma_k}$, $\rho''=N_k''/V_{\Gamma_k}$ and
$\rho=(\rho'+\rho'')/2$.  By linear interpolation Eq.~(\ref{convex})
holds for any $\rho'$ and $\rho''$.  Taking the limit $k \rightarrow
\infty$ gives
\be
\label{convex2}
\la f_\Gamma(\rho) \ra\le{1\over 2}\(
\la f_\Gamma(\rho')\ra+\la f_\Gamma(\rho'')\ra\).
\ee
Therefore, $\langle f_\Gamma(\rho)\rangle\equiv{\bar f_\Gamma(\rho)}$ 
is a convex function of $\rho$. Since ${\bar f_\Gamma(\rho)}$ is
a bounded function of $\rho$, it is also continuous function 
of $\rho$~\cite{hardy}.

\newpage

\begin{figure}[tb]
\vspace{0.3in}
\label{distribution}
\centerline{\psfig{figure=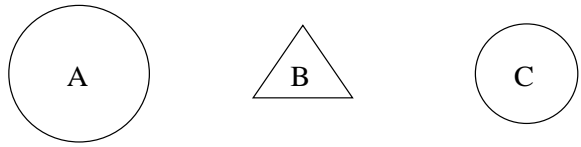,width=3in}}
\vspace{0.3in}
\caption{ An example of a size and distribution of particles. The hard 
core particles A,
B, C are chosen with fixed probabilities.}
\end{figure}

\begin{figure}[tb]
\vspace{0.3in}
\label{choice}
\centerline{\psfig{figure=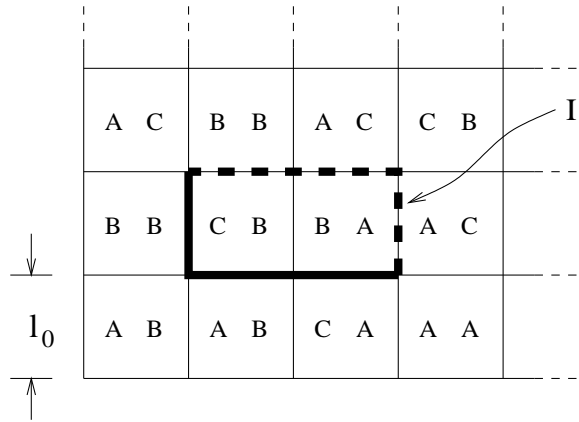,width=3in}}
\vspace{0.3in}
\caption {One particular choice of particles from the distribution in Fig 1.
into basic cubes, with $N_0=2$ particles in each basic cube. 
The box $I=${\bf [}$(l_0,l_0),(3l_0,2l_0)${\bf )}
in this example is assigned particles C, B, B, and A. For easy
visualization we show the figure in two-dimensions instead of the actual 
case of three dimensions.}
\end{figure}

\begin{figure}[tb]
\label{anyshape}
\centerline{\psfig{figure=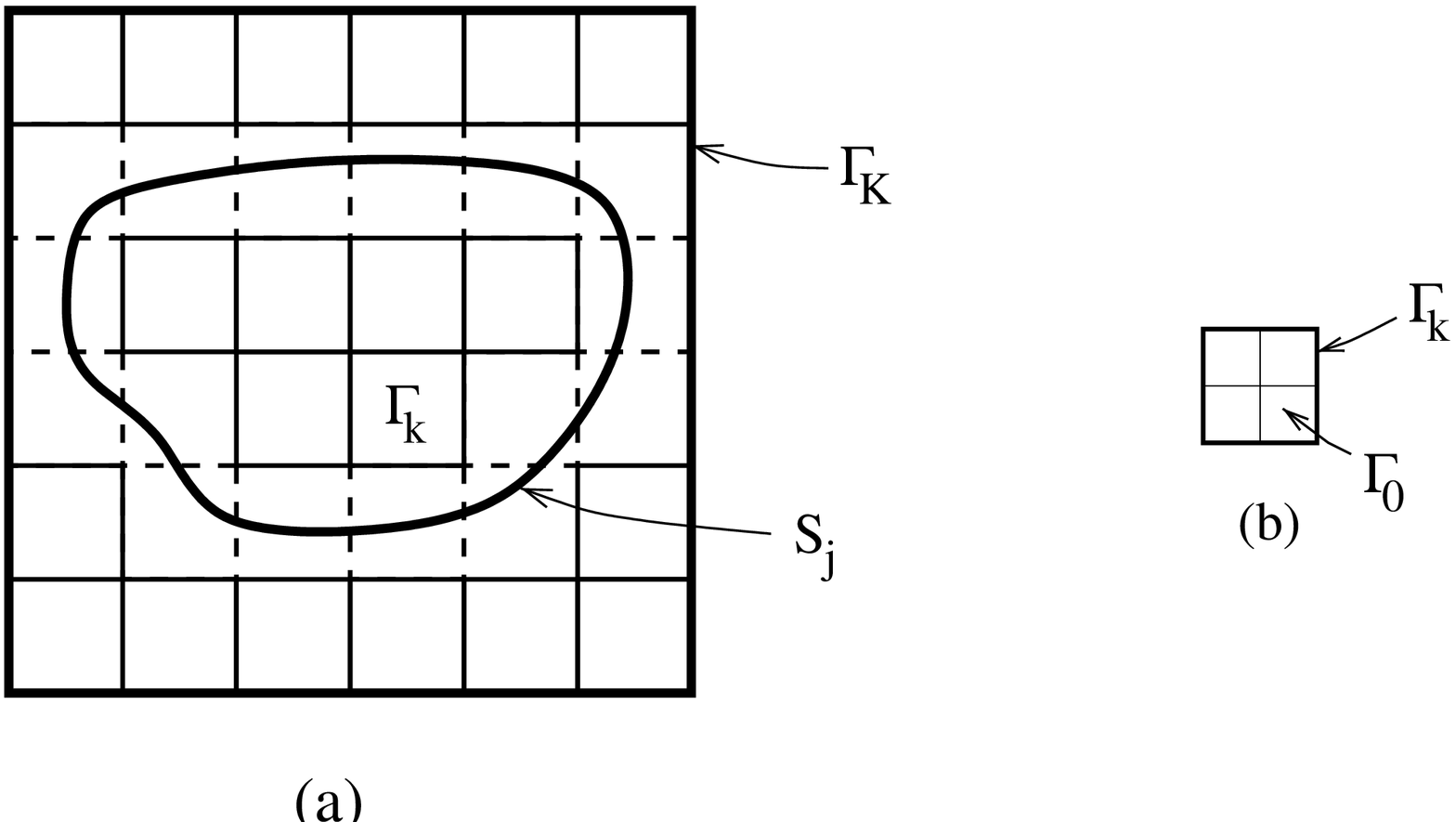,width=6in}}
\vspace{0.2in}
\caption{(a) A cross-section of an arbitrary shape $S_j$. The space 
inside $S_j$, and the
space between $S_j$ and $\Gamma_K$, is maximally filled by cubes 
$\Gamma_k$. Dashed lines indicate
the cubes that intersect the boundary of $S_j$.  The volume of 
cubes $\Gamma_k$ that intersect $S_j$ becomes negligible
in comparison to the volume of $S_j$ as the size of $S_j$ goes 
to infinity and the size of $\Gamma_k$ 
grows less rapidly than $S_j$. (b) A cube $\Gamma_k$ 
in $d$ dimensions consists of $2^{kd}$ basic cubes $\Gamma_0$. 
In this case $k=1$ and $d=2$.}
\end{figure}

\begin{figure}[tb]
\label{particles}
\centerline{\psfig{figure=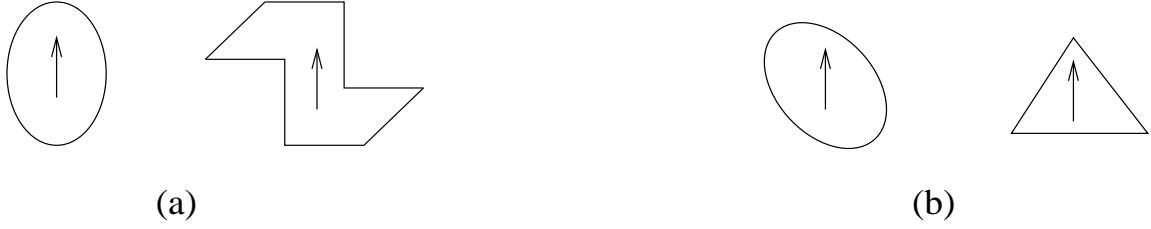,width=6in}}
\vspace{0.4in}
\caption{(a) Particle shapes with a two-fold symmetry perpendicular 
to their magnetization. The 
arrow indicates the direction of magnetization. (b) Particle shapes 
lacking two-fold 
symmetry perpendicular to their magnetization.}
\end{figure}

\end{document}